\documentclass[12pt]{article}
\usepackage{epsfig}
\begin{document}
\title{Non-resonant kaon pair production and medium effects in proton--nucleus collisions}
\author{E. Ya. Paryev$^{1,2}$, M. Hartmann$^3$, Yu. T. Kiselev$^2$\\
{\it $^1$Institute for Nuclear Research, Russian Academy of Sciences,}\\
{\it Moscow 117312, Russia}\\
{\it $^2$Institute for Theoretical and Experimental Physics,}\\
{\it Moscow 117218, Russia}\\
{\it $^3$Institut f$\ddot{u}$r Kernphysik and J$\ddot{u}$lich Centre for Hadron Physics,}\\
{\it Forschungszentrum J$\ddot{u}$lich, D--52425 J$\ddot{u}$lich, Germany}}

\renewcommand{\today}{}
\maketitle

\begin{abstract}
   We study the non-resonant (non-$\phi$) production of $K^+K^-$ pairs by protons of 2.83 GeV kinetic
   energy on C, Cu, Ag, and Au targets within the collision model, based on the nuclear spectral function,
   for incoherent primary proton--nucleon and secondary pion--nucleon creation processes.
   The model takes into account the initial proton and final kaon absorption, target nucleon
   binding and Fermi motion as well as nuclear mean-field potential effects on these processes.
   We calculate the antikaon momentum dependences of the exclusive absolute and
   relative $K^+K^-$ pair yields in the acceptance window of the ANKE magnetic spectrometer, used in a
   recent experiment performed at COSY, within the different scenarios for the antikaon-nucleus optical
   potential.
   We demonstrate that the above observables are strongly sensitive to this potential.
   Therefore, they can be useful to help determine the $K^-$ optical potential from the
   direct comparison of the results of our calculations with the data from the respective ANKE-at-COSY experiment. We also show that the pion--nucleon production channels dominate in the low-momentum
   $K^-$, $K^+$ production in the considered kinematics and, hence, they have to be accounted for in the analysis
   of these data.
\end{abstract}

\newpage

\section*{1. Introduction}

\hspace{1.5cm} The study of kaon and antikaon properties in a strongly interacting environment was a very
active research field over the last two decades (see, for example, [1--3]), especially in connection with
the questions of the partial restoration of chiral symmetry in hot/dense nuclear matter and of the existence
of the $K^-$ condensate in neutron stars. From the analysis of KaoS [4], FOPI [5] and ANKE [6] data on $K^+$
production in heavy--ion and proton--nucleus reactions in the framework of transport approaches [7] and [8]
was established that the $K^+$ meson ($K^+={\bar s}u$) feels a moderately repulsive nuclear potential of
about 20--30 MeV at normal nuclear matter density $\rho_0$, in agreement with theoretical calculations [3, 9].
A very recent results on the azimuthal emission pattern of $K^+$ mesons in Ni$+$Ni collisions at a beam kinetic
energy of 1.91 A GeV, obtained by the FOPI Collaboration, also support the existence of a repulsive
kaon--nucleus potential of the same order of magnitude as that mentioned above [10]. Similar to $K^+$,
the $K^0$ meson in-medium nuclear potential is expected to be repulsive as well due to its quark content
($K^0={\bar s}d$). Indeed, the comparison of the ratio of the $K^0$ momentum distributions from
${\pi^-}+$Pb and ${\pi^-}+$C reactions at a pion incident momentum of 1.15 GeV/c, measured by the FOPI
Collaboration, with the HSD transport model calculations suggests a repulsive $K^0$--nucleus optical potential
of about 20 MeV at normal nuclear matter density [11]. Whereas the data on $K^0$ transverse momentum spectra
in Ar$+$KCl and $p+$Nb reactions at bombarding kinetic energies of 1.756 A GeV and 3.5 GeV,
collected recently by the HADES Collaboration,
point to the existence of a stronger repulsive in-medium $K^0$ potential of about 35--40 MeV strength
at saturation density $\rho_0$ for kaons at rest [12].

  Contrary to the $K^+$, $K^0$ mesons, the $K^-$ meson ($K^-=s{\bar u}$) properties in nuclear matter
are much less known nowadays
and are still very intense debated. This is due to the fact that antikaon can easily form baryonic resonances in
the nuclear medium, which leads to the modification of its in-medium properties and requires a complicated
self-consistent coupled-channel calculations of this modification with including the complete
set of the pseudoscalar meson and baryon octets. Such calculations, based on a chiral Lagrangians [13--19] or on
meson-exchange potentials [20, 21], predict a relatively shallow low-energy $K^-$--nucleus potential with a central
depth of the order of -50 to -80 MeV. On the other hand, fits of the $K^-$ atomic data [22] by the phenomenological
density-dependent optical potentials as well as by these potentials and the relativistic mean-field calculations [23]
lead to a deep potential of depth of about -200 MeV at density $\rho_0$ for antikaon in matter, which is in conflict
with the results of the self-consistent approaches mentioned above. Recent a chirally motivated meson-baryon
coupled-channel model [24], which accounts for the subthreshold in-medium $K^-N$ $s$-wave scattering amplitudes,
produces $K^-$ potential depths in the range of -(80--90) MeV in kaonic atoms at nuclear
matter density. But by adding a ${\rho}^2$-dependent phenomenological term, which could represent
${\bar K}NN \to YN$ absorption and dispersion, to improve the agreement with data, the antikaon potential becomes
twice as deep [24, 25]. However, it should be noted that the antikaonic-atom data probe medium at the surface of the
nucleus and, therefore, do not provide the strong constraints on the $K^-$--nucleus potential at the normal nuclear
matter density. Present status of the study of low-energy in-medium $K^-$ nuclear interactions is given in [25].

   Motivated by the fact that a very strong antikaon--nucleon potential could form a deeply bound kaonic states [26],
the experiments of KEK [27] and FINUDA [28] have been performed to search for such states and, as a result, they claim
evidence for their existence. In DISTO experiment [29], the indication of a deeply bound compact $K^-pp$ cluster formed in the $pp \to p{\Lambda}K^+$ reaction at 2.85 GeV has been obtained. An analysis of the final state of this reaction
at a beam kinetic energy of 3.5 GeV with the aim of studying of the possible formation of an intermediate $K^-pp$
bound-state in it has been carried out by the HADES Collaboration as well [30]. It was found that the data,
which were collected with the HADES spectrometer, are compatible with the background-only hypothesis that includes
no production of intermediate kaonic cluster. Also, more precise KEK experiment [31]
did not observe a narrow structure in proton spectrum following $K^-$ absorption at rest in $^4$He, which has been
identified in [27] with deeply bound $K^-ppn$ cluster. In a recent J-PARC experiment [32], devoted to searching
for the $K^-pp$ bound state in the in-flight $^3$He($K^-,n$) reaction at 1 GeV/c, no significant peak was also observed in the neutron missing-mass spectrum in the region corresponding to $K^-pp$ binding energy larger than 80 MeV, where
a bump structure was reported in the ${\Lambda}p$ final state by the FINUDA and DISTO Collaborations.
On the other hand, an another very recent J-PARC experiment [33] claims evidence for the observation of a
"$K^-pp$"-like structure in the $d(\pi^+,K^+)$ reaction at 1.69 GeV/c with binding energy of
$95^{+18}_{-17}$(stat.)$^{+30}_{-21}$(syst.) MeV. Theoretically, it has
been argued [34] that at present there are no experimental evidences both for the existence of deeply
bound kaonic states and for a strong antikaon--nucleus potential with a central depth of the order of -200 MeV, claimed in the experiment [35].

    The information about in-medium properties of antikaons can be deduced also from the study of their production
both in heavy--ion and proton-nucleus collisions at incident energies near or below the free nucleon--nucleon
threshold (2.5 GeV) because the dropping $K^-$ mass scenario will lead to an enhancement of the $K^-$ yield in
these collisions due to in-medium shifts of the elementary production thresholds to lower energies.
Thus, in particular, in [1, 3, 10] was shown that the existence of a $K^-$ condensate is not
compatible with the available heavy--ion data. Analysis of the KaoS data on the ratio of $K^-$ and $K^+$ inclusive
momentum spectra from reactions $p+A \to K^{\pm}+X$ with $A=$C and Au at laboratory angles from 36$^{\circ}$ to
60$^{\circ}$ and beam energy of 2.5 GeV within the BUU transport model has shown that these data are consistent with
the in-medium $K^-A$ potential of the order of -80 MeV at normal nuclear density [36]. Reasonable description of the
measured at ITEP accelerator both inclusive antikaon momentum distributions in the momentum range from 0.6 to
1.3 GeV/c at a lab angle of 10.5$^{\circ}$ in $p$Be, $p$Cu interactions at 2.25, 2.4 GeV beam energies and $K^-$
excitation functions in these interactions for $K^-$ momentum of 1.28 GeV/c at bombarding energies $<$ 3 GeV was,
respectively, reached in [37] and in [38] in the frame of folding model, based on the target nucleon momentum
distribution and on free elementary cross sections, assuming vacuum $K^+$, $K^-$ masses. The $K^-$ potential of
$\approx$ -28 MeV at density $\rho_0$ and an antikaon momentum of 800 MeV/c has been extracted in [39] from the
data on elastic $K^-A$ scattering within Glauber theory. So, one can conclude that the situation with the
antikaon--nucleus optical potential is still very unclear.

   To make progress in understanding the strength of the $K^-$ interaction with nuclear medium, it is necessary
to carry out the exclusive measurements with the tagging the low-momentum $K^-$ mesons not stemming from the
vacuum $\phi$ decays and, therefore, bringing the "genuine" information about this strength. Such
measurements have been recently performed by the ANKE Collaboration at COSY that studied the production of
$K^+K^-$ pairs in proton collisions with C, Cu, Ag, and Au targets at an incident beam energy of 2.83 GeV for
the $K^+K^-$ invariant masses belonging to the $\phi$ and non-$\phi$ regions [40, 41]. For the first time, in the
$\phi$ region, the momentum dependence of the $\phi$ nuclear transparency ratio, the in-medium $\phi$ meson width,
and the differential cross section for its production at forward angles have been determined for these targets over
the $\phi$ momentum range of 0.6--1.6 GeV/c [40, 41]. For the purpose of obtaining of the exclusive differential
cross section for $K^+K^-$ pair production on the considered target nuclei as a function of the laboratory $K^-$
momentum, a further analysis of the data coming from the non-$\phi$ region of the $K^+K^-$ invariant mass is
currently being carried out.

In this respect, the main purpose of the present work is to get the estimates of the absolute and relative yields of
non-resonant (not through the vacuum $\phi \to K^+K^-$ decays
\footnote{$^)$Since the $\phi$ mesons produced in $pA$ reactions at incident energy of 2.83 GeV have
large momenta [41], most of them will decay outside the nucleus. Moreover, the total cross section of
the reaction $pp \to pp{\phi} \to ppK^+K^-$, as is seen from figure 1 given below, is more than one order of
magnitude less than that of the non-resonant $K^+K^-$ pair production in $pp$ collisions at excess
energies $\epsilon_{K{\bar K}}$ above the $K^+K^-pp$ threshold $\sqrt{s_{\rm th}}$,
$\epsilon_{K{\bar K}}=\sqrt{s}-\sqrt{s_{\rm th}}$ ($\sqrt{s}$ is the total center-of-mass energy),
falling in the range of 0.1--0.5 GeV, which mainly contribute, as our estimates showed, to the $K^+K^-$
creation in the considered in the next section primary processes at this incident energy.
Therefore, we can neglect the possible
contribution of the distorted due to the FSI $K^+K^-$ pairs stemming from the in-medium
$\phi \to K^+K^-$ decays to the "genuine" non-resonant kaon pair production.}$^)$)
$K^+K^-$ pairs from $pA$ collisions
at beam energy of 2.83 GeV in the acceptance window of the ANKE spectrometer used in the experiment [40, 41]
performed at COSY accelerator in the framework of the collision model, based on the nuclear spectral function,
for incoherent primary proton--nucleon and secondary pion--nucleon $K^+K^-$ creation processes in different
scenarios for the $K^-$ nuclear potential. In view of the expected data from this
experiment, the estimates can be used as an important tool for determining the antikaon-nucleus optical potential.

\section*{2. The model}
\section*{2.1. Direct non-resonant $K^+K^-$ pair production mechanism}

\hspace{1.5cm} The direct non-resonant (non-$\phi$) production of $K^+K^-$ pairs in $pA$ reactions
at initial energy of 2.83 GeV of our interest can occur due to the nucleon's Fermi motion
in the following elementary processes with zero, one and two pions in the final states
\footnote{$^)$It should be pointed out that the $K^+K^-$ pairs can be produced also in the elementary
reaction channel non-$\phi$ $pn \to dK^+K^-$ studied recently in the threshold energy region by the
ANKE Collaboration [42]. However, its contribution to the kaon pair creation in $pA$ collisions
at beam energy of 2.83 GeV is expected [37] to be insignificant at antikaon momenta $<$ 0.9 GeV/c
of our interest.}$^)$
:
\begin{equation}
p+p \to K^++K^-+p+p,
\end{equation}
\begin{equation}
p+p \to K^++K^-+N+N+m\pi;
\end{equation}
\begin{equation}
p+n \to K^++K^-+p+n,
\end{equation}
\begin{equation}
p+n \to K^++K^-+N+N+m\pi,
\end{equation}
where $m=1, 2$.
In the following calculations we will include the medium modification of the final nucleons, kaon and antikaon, participating in the production processes (1)--(4), by using
in the in-medium cross sections of these processes, for reasons of numerical simplicity,
instead of their local effective masses $m^*_{h}(r)$ their average in-medium masses $<m^*_{h}>$
defined in line with [43, 44, 45] as:
\begin{equation}
<m^*_{h}>=m_{h}+U_h^0\frac{<{\rho_N}>}{{\rho_0}}+V_{ch}(R_{1/2}).
\end{equation}
Here, $m_{h}$ is the hadron free space mass
\footnote{$^)$The bare kaon and antikaon masses are denoted by $m_K$.}$^)$,
$U_h^0$ is its nuclear potential depth at saturation
density ${\rho_0}$, $<{\rho_N}>$ is the average nucleon density
\footnote{$^)$Which is assumed to be equal to ${\rho_0}/2$ [46] for all considered target
nuclei.}$^)$
and $V_{ch}(R_{1/2})$ is the charged hadron Coulomb potential
\footnote{$^)$It is worth mentioning that this potential for kaons and protons amounts approximately
to 3.5, 9.7, 12.3, and 16.6 MeV for C, Cu, Ag, and Au nuclei, respectively [44], i.e. for heavy
nuclei like Au it is of the same order of magnitude as the repulsive $K^+$ nuclear
mean-field potential.}$^)$
taken at the point $R_{1/2}$ at which the local nucleon density ${\rho_N(r)}$ satisfies the
relation $\rho_N(R_{1/2})=\rho_0/2$. In the subsequent study for the $K^+$ and $K^-$ mass shifts
$U_{K^+}^0$ and $U_{K^-}^0$ we will employ the following options: $U_{K^+}^0=22$ MeV [43] and
i) $U_{K^-}^0=0$, ii) $U_{K^-}^0=-60$ MeV, iii) $U_{K^-}^0=-126$ MeV [43], and iv) $U_{K^-}^0=-180$ MeV.
One can see that the second option corresponds to the shallow antikaon-nucleus potential, whereas the
third and fourth ones represent, respectively, the relatively deep and deep $K^-A$ optical potentials.
Accounting for that the relation between the effective scalar nucleon potential $U_N^0$, entering into
the equation (5), and the corresponding Schr${\ddot{\rm o}}$dinger equivalent potential $V_{NA}^{\rm SEP}$ at the
normal nuclear matter density  is given by
\begin{equation}
U_N^0=\frac{\sqrt{m_N^2+p_N^2}}{m_N}V_{NA}^{\rm SEP}
\end{equation}
as well as the fact that the momenta $p_N$ of the outgoing in reactions (1)--(4) nucleons are,
as showed our calculations, around of momentum of 1 GeV/c and using that
$V_{NA}^{\rm SEP} \approx 25$ MeV at this momentum [47], we can readily obtain that $U_N^0 \approx 35$ MeV.
We will employ this potential throughout our present work.
The total energy $E^\prime_{h}$ of the hadron inside the nuclear medium can be
expressed through its average effective mass $<m^*_{h}>$ defined above and its in-medium momentum
${\bf p}^{\prime}_{h}$ as in the free particle case, namely:
\begin{equation}
E^\prime_{h}=\sqrt{(<m^*_{h}>)^2+({\bf p}^{\prime}_{h})^2}.
\end{equation}
The momentum ${\bf p}^{\prime}_{h}$ is related to the vacuum one ${\bf p}_{h}$
by the following expression:
\begin{equation}
E^\prime_{h}=\sqrt{(<m^*_{h}>)^2+({\bf p}^{\prime}_{h})^2}=
\sqrt{m^2_{h}+{\bf p}^2_{h}}=E_h,
\end{equation}
where $E_h$ is the hadron vacuum total energy.

  Then, taking into consideration the distortions of the initial proton and final kaons as well as
the fact that in the ANKE experiment the latter were detected in the forward polar angle domain
$0^{\circ} \le \theta_{K^+,K^-} \le 12^{\circ}$ and using the results given in [37, 43--45], we can represent
the exclusive differential cross section for the production of $K^+$ meson with the vacuum momentum
${\bf p}_{K^+}$ in coincidence with the $K^-$ meson with the vacuum momentum ${\bf p}_{K^-}$ off nuclei
in the primary proton-induced reaction channels (1)--(4) as follows:
\begin{equation}
\frac{d\sigma_{pA\to K^+K^-X}^{({\rm prim})}
({\bf p}_0,{\bf p}_{K^+},{\bf p}_{K^-})}
{d{\bf p}_{K^+}d{\bf p}_{K^-}}=I_{V}[A]
\end{equation}
$$
\times
\left[\frac{Z}{A}
\left<\frac{d\sigma^{\rm nr}_{pp\to K^+K^-X}({\bf p}^{\prime}_{0},
{\bf p}^{\prime}_{K^+},{\bf p}^{\prime}_{K^-})}{d{\bf p}^{\prime}_{K^+}d{\bf p}^{\prime}_{K^-}}\right>_A+
\frac{N}{A}
\left<\frac{d\sigma^{\rm nr}_{pn\to K^+K^-X}({\bf p}^{\prime}_{0},
{\bf p}^{\prime}_{K^+},{\bf p}^{\prime}_{K^-})}{d{\bf p}^{\prime}_{K^+}d{\bf p}^{\prime}_{K^-}}\right>_A
\right]
\frac{d{\bf p}^{\prime}_{K^+}}{d{\bf p}_{K^+}}
\frac{d{\bf p}^{\prime}_{K^-}}{d{\bf p}_{K^-}},
$$
where
\begin{equation}
I_{V}[A]=2{\pi}A\int\limits_{0}^{R}r_{\bot}dr_{\bot}
\int\limits_{-\sqrt{R^2-r_{\bot}^2}}^{\sqrt{R^2-r_{\bot}^2}}dz
\rho(\sqrt{r_{\bot}^2+z^2})
\end{equation}
$$
\times
\exp{\left[-\sigma_{pN}^{\rm in}A\int\limits_{-\sqrt{R^2-r_{\bot}^2}}^{z}\rho(\sqrt{r_{\bot}^2+x^2})dx
-\sigma_{K^+N}^{\rm tot}A\int\limits_{z}^{\sqrt{R^2-r_{\bot}^2}}
\rho(\sqrt{r_{\bot}^2+x^2})dx\right]}
$$
$$
\times
\exp{\left[-\int\limits_{z}^{\sqrt{R^2-r_{\bot}^2}}
\mu_{K^-N}\left[p^{\prime}_{K^-}(\sqrt{r_{\bot}^2+x^2})\right]
\rho(\sqrt{r_{\bot}^2+x^2})dx\right]},
$$
\begin{equation}
\left<\frac{d\sigma^{\rm nr}_{pN\to K^+K^-X}({\bf p}^{\prime}_{0},
{\bf p}^{\prime}_{K^+},{\bf p}^{\prime}_{K^-})}{d{\bf p}^{\prime}_{K^+}d{\bf p}^{\prime}_{K^-}}\right>_A
=
\int\int
P_A({\bf p}_t,E)d{\bf p}_tdE
\end{equation}
$$
\times
\left\{\frac{d\sigma^{\rm nr}_{pN\to K^+K^-X}[\sqrt{s},<m^*_{K^+}>,<m^*_{K^-}>,<m^*_{N}>,<m^*_{N}>,
{\bf p}^{\prime}_{K^+},{\bf p}^{\prime}_{K^-}]}{d{\bf p}^{\prime}_{K^+}d{\bf p}^{\prime}_{K^-}}\right\}
$$
and
\begin{equation}
\mu_{K^-N}\left[p^{\prime}_{K^-}(r^{\prime})\right]=
\sigma^{\rm tot}_{K^-p}\left[p^{\prime}_{K^-}(r^{\prime})\right]Z+
\sigma^{\rm tot}_{K^-n}\left[p^{\prime}_{K^-}(r^{\prime})\right]N,
\end{equation}
\begin{equation}
p^{\prime}_{K^-}(r^{\prime})=\sqrt{E^2_{K^-}-[m^*_{K^-}(r^{\prime})]^2},
\end{equation}
\begin{equation}
m^*_{K^-}(r^{\prime})=m_{K}+U_{K^-}^0\frac{\rho_N(r^{\prime})}{{\rho_0}}+V_{cK^-}(r^{\prime}),
\end{equation}
\begin{equation}
r^{\prime}=\sqrt{r_{\bot}^2+x^2}.
\end{equation}
Here,
$d\sigma^{\rm nr}_{pN\to K^+K^-X}[\sqrt{s},<m^*_{K^+}>,<m^*_{K^-}>,<m^*_{N}>,<m^*_{N}>,
{\bf p}^{\prime}_{K^+},{\bf p}^{\prime}_{K^-}]/d{\bf p}^{\prime}_{K^+}d{\bf p}^{\prime}_{K^-}$
are the "in-medium" differential cross sections for the non-resonant production of $K^+$ and $K^-$
mesons with the
in-medium momenta ${\bf p}^{\prime}_{K^+}$ and ${\bf p}^{\prime}_{K^-}$, respectively, in reactions
(1) and (2) ($N=p$) as well as in (3) and (4) ($N=n$) at the $pN$ center-of-mass energy $\sqrt{s}$
\footnote{$^)$The expression for $s$ is given in [43] by the formula (18).}$^)$
;
$\rho(r)$ and $P_A({\bf p}_t,E)$ are the local nucleon density and the
spectral function of target nucleus $A$ normalized to unity
\footnote{$^)$The specific information about used in our subsequent calculations these quantities
is given in [43--45, 48].}$^)$
;
${\bf p}_{t}$  and $E$ are the internal momentum and binding energy of the struck target nucleon
just before the collision; $\sigma_{pN}^{\rm in}$ and $\sigma_{K^+N}^{\rm tot}$, $\sigma_{K^-N}^{\rm tot}$
are the inelastic and total cross sections
of the free $pN$ and $K^+N$, $K^-N$ interactions
\footnote{$^)$We use $\sigma_{pN}^{\rm in}=30$ mb for the considered projectile proton energy and
$\sigma_{K^+N}^{\rm tot}=12$ mb for all kaon momenta of interest [37] in our calculations.
For the antikaon-nucleon
total cross sections $\sigma_{K^-p}^{\rm tot}(p_{K^-})$, $\sigma_{K^-n}^{\rm tot}(p_{K^-})$
as functions of the $K^-$ momentum $p_{K^-}$ we employ in them the corresponding parametrizations
suggested in [49] (see, also, [37]). Dealing with the total cross sections
$\sigma_{K^{\pm}N}^{\rm tot}$ in (10), we assume that if a kaon undergoes a quasi-elastic collision
with the intranuclear nucleon it will not fall in the ANKE acceptance window.}$^)$
;
$Z$ and $N$ are the numbers of protons and neutrons in
the target nucleus ($A=Z+N$), $R$ is its radius;
${\bf p}_{0}$ and ${\bf p}^{\prime}_{0}$ are the momenta of the initial proton outside and inside
the target nucleus
\footnote{$^)$They are linked by the equation (13) from [43].}$^)$
.
The quantity $I_{V}[A]$ in equation (9) represents the effective number of target
nucleons participating in the primary $pN\to K^+K^-X$ reactions. It is determined, in particular,
by the one nucleon $K^-$ absorption and does not account for the two nucleon antikaon absorption
mechanism discussed in [46]. The inclusion of the $K^-$ absorption by pairs of nucleons of 20\% that
of the one body absorption in line with [46] leads to only small corrections to the quantity $I_{V}[A]$
\footnote{$^)$And to the quantity $I_{V}^{\rm sec}[A]$, determining the differential cross section
for non-resonant $K^+K^-$ pair production in $pA$ reactions from the two-step creation
mechanism (see below), as our calculations also showed.}$^)$
.
They are within 1--3\% for gold nucleus and for the antikaon momenta of interest, as our calculations
for this nucleus with the diffuse boundary showed. Therefore, we ignored the two nucleon $K^-$ absorption
mechanism in the present study.

 In our calculations of the non-resonant $K^+K^-$ pair creation in $pA$ interactions reported below
the in-medium exclusive differential cross sections
$d\sigma^{\rm nr}_{pN\to K^+K^-X}[\sqrt{s},<m^*_{K^+}>,<m^*_{K^-}>,<m^*_{N}>,<m^*_{N}>,
{\bf p}^{\prime}_{K^+},{\bf p}^{\prime}_{K^-}]/d{\bf p}^{\prime}_{K^+}d{\bf p}^{\prime}_{K^-}$
have been described according to the four-body phase space (see, also, [43])
\footnote{$^)$It should be noted that this space was found [50--52] to be distorted mostly by the
$pp$ and $K^-p$ FSI in the non-resonant $pp \to ppK^+K^-$ reaction both at beam energy of 2.83 GeV
and at incident energies below the $\phi$ meson production threshold. However, since we are interested in
the exclusive cross sections for $K^+K^-$ production in $pA$ reactions averaged over the ANKE
acceptance polar angle domain $0^{\circ} \le \theta_{K^+,K^-} \le 12^{\circ}$ (see below), but not
in the full differential ones, one may hope that the account of this distortion as well as the deviation
of the differential cross sections for the $K^+K^-NN$ production in reactions (2) and (4) from those
dictated by the four-body phase space leads to insignificant corrections.}$^)$
:
\begin{equation}
\frac{d\sigma^{\rm nr}_{pN\to K^+K^-X}[\sqrt{s},<m^*_{K^+}>,<m^*_{K^-}>,<m^*_{N}>,<m^*_{N}>,
{\bf p}^{\prime}_{K^+},{\bf p}^{\prime}_{K^-}]}{d{\bf p}^{\prime}_{K^+}d{\bf p}^{\prime}_{K^-}}=
\frac{\sigma^{\rm nr}_{pN\to K^+K^-X}(\sqrt{s},\sqrt{s^*_{\rm th}})}{4E^{\prime}_{K^+}E^{\prime}_{K^-}}
\end{equation}
$$
\times
\frac{I_2(s_2,<m^*_N>,<m^*_N>)}{I_4(s,<m^*_{K^+}>,<m^*_{K^-}>,<m^*_N>,<m^*_N>)},
$$
where
\begin{equation}
I_2(s_2,<m^*_N>,<m^*_{N}>)=\left(\frac{\pi}{2}\right)\frac{\lambda[s_2,(<m^*_N>)^{2},(<m^*_N>)^2]}{s_2},
\end{equation}
\begin{equation}
s_2=s+s^{\prime}_{K^+K^-}-2(E^{\prime}_0+E_t)(E^{\prime}_{K^+}+E^{\prime}_{K^-})+
2({\bf p}^{\prime}_{0}+{\bf p}_t)({\bf p}^{\prime}_{K^+}+{\bf p}^{\prime}_{K^-}),
\end{equation}
\begin{equation}
s^{\prime}_{K^+K^-}=(E^{\prime}_{K^+}+E^{\prime}_{K^-})^2-
({\bf p}^{\prime}_{K^+}+{\bf p}^{\prime}_{K^-})^2
\end{equation}
and the quantities $E^{\prime}_0$, $E_t$, $I_4$ and $\lambda$ are defined in [43] by the
equations (12), (15), (28) and (29), respectively.
Here, $\sigma^{\rm nr}_{pN\to K^+K^-X}(\sqrt{s},\sqrt{s^*_{\rm th}})$ are the "in-medium"
total cross sections for the non-resonant $K^+K^-$ pair production in reactions (1) and (2)
($N=p$) as well as in (3) and (4) ($N=n$) having the
threshold energy $\sqrt{s^*_{\rm th}}=<m^*_{K^+}>+<m^*_{K^-}>+2<m^*_N>$. Following [43], we assume
that these cross sections are equivalent to the vacuum ones
$\sigma^{\rm nr}_{pN\to K^+K^-X}(\sqrt{s},\sqrt{s_{\rm th}})$
in which the free threshold energy $\sqrt{s_{\rm th}}=2(m_K+m_N)$
is replaced by the in-medium threshold one $\sqrt{s^*_{\rm th}}$.

 Let us now specify the total cross sections $\sigma^{\rm nr}_{pp\to K^+K^-X}(\sqrt{s},\sqrt{s_{\rm th}})$
and $\sigma^{\rm nr}_{pn\to K^+K^-X}(\sqrt{s},\sqrt{s_{\rm th}})$.
Due to the lack of experimental information about the non-resonant production of $K^+K^-$ pairs in
$pp$ interactions (1) and (2) at excess energies above the $\pi$ meson mass $m_{\pi}$, the first cross
section can be defined as:
\begin{equation}
\sigma^{\rm nr}_{pp\to K^+K^-X}(\sqrt{s},\sqrt{s_{\rm th}})=\sigma_{pp\to K^-X}(\sqrt{s},\sqrt{s_{\rm th}})-
\sigma_{pp\to pp\phi}(\sqrt{s})BR(\phi \to K^+K^-)
\end{equation}
$$
-\sigma_{pp\to K^0K^-pp\pi^+}(\sqrt{s})-\sigma_{pp\to K^0K^-pn\pi^+\pi^+}(\sqrt{s})-
\sigma_{pp\to K^0K^-pp\pi^0\pi^+}(\sqrt{s}),
$$
where $\sigma_{pp\to K^-X}(\sqrt{s},\sqrt{s_{\rm th}})$, $\sigma_{pp\to pp\phi}(\sqrt{s})$,
$\sigma_{pp\to K^0K^-pp\pi^+}(\sqrt{s})$, $\sigma_{pp\to K^0K^-pn\pi^+\pi^+}(\sqrt{s})$ and \\
$\sigma_{pp\to K^0K^-pp\pi^0\pi^+}(\sqrt{s})$ are the total cross sections of the reactins
$pp \to K^-X$, $pp \to pp{\phi}$, $pp \to K^0K^-pp\pi^+$, $pp \to K^0K^-pn\pi^+\pi^+$ and
$pp \to K^0K^-pp\pi^0\pi^+$, respectively, whereas
$BR(\phi \to K^+K^-)=0.491$
\footnote{$^)$We neglect the possible modification of the branching ratio of the decay
$\phi \to K^+K^-$ in the nuclear medium.}$^)$
.
The available experimental data for the three latter channels are quite scarse and they were
approximated in [53] by formulas (63), (64) and (67), which we adopt for our present study.
The results of calculations of the total cross section of the first channel and the sum of the
total cross sections of the second and third ones, according to them, are shown in
figure 1 by dotted and dash-dot-dotted lines, respectively.
\begin{figure}[!h]
\begin{center}
\includegraphics[width=12.0cm]{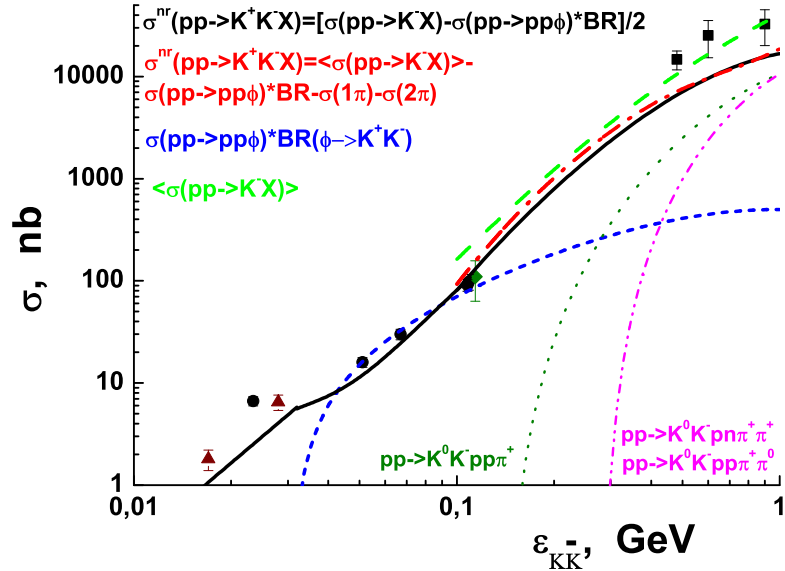}
\vspace*{-2mm} \caption{(color online) The kaon and kaon-antikaon production total cross sections
in proton--proton collisions as functions of the excess energy above the $K^+K^-pp$ threshold.
For notation see the text.}
\label{void}
\end{center}
\end{figure}
For the total cross section $\sigma_{pp\to pp\phi}(\sqrt{s})$
we have used in it parametrization (12) from [48], which
was multiplied by the factor of 0.75 in the light of the improved data on the $\phi$
production total cross section obtained in [50, 51]. Multiplying this cross section on the
branching ratio of the $\phi$ decay into $K^+K^-$ pair, we get the total cross section of the
reaction $pp\to pp\phi \to ppK^+K^-$, which is shown in figure 1 by short-dashed line.
Up to now, there have been no data on the inclusive cross section
$\sigma_{pp\to K^-X}(\sqrt{s},\sqrt{s_{\rm th}})$ at excess energies
$m_{\pi} < \epsilon_{K{\bar K}} < 0.5$ GeV of our interest and the existing parametrizations
(31) from [43] and (4) from [54] give here different predictions for this cross section
(see figure 1 in [43])
\footnote{$^)$Their arithmetic average is depicted in figure 1 by long-dashed curve,
which is well in line with the inclusive $pp \to K^-X$ data (full squares) at higher energies
taken from [43].}$^)$
.
Due to this fact, to reduce the uncertainty in determination of the cross
section (20) it is natural to estimate it by using both of these parametrizations and for this cross
section to take an arithmetic average of the obtained results. Such procedure has been carried out and
after it was found that the above average, presented in figure 1 by dot-dashed curve,
can be well fitted by the following expression:
\begin{equation}
\sigma^{\rm nr}_{pp\to K^+K^-X}(\sqrt{s},\sqrt{s_{\rm th}})=\frac{1}{2}
\left[\sigma_{pp\to K^-X}(\sqrt{s},\sqrt{s_{\rm th}})-
\sigma_{pp\to pp\phi}(\sqrt{s})BR(\phi \to K^+K^-)\right],
\end{equation}
where inclusive cross section $\sigma_{pp\to K^-X}(\sqrt{s},\sqrt{s_{\rm th}})$ is taken from [43]
in parametrization (31) (solid line in figure 1).
As can be seen from figure 1, this expression reproduces reasonable well also
the available experimental data on the non-$\phi$ contribution to the $pp \to K^+K^-pp$ total cross
section at low excess energies from DISTO [55] (full diamond),
COSY-11 [56] (full triangles) and ANKE [50--52] (full circles) Collaborations.
Since there are no data for $K^+K^-$ production in $pn$ reactions (3), (4), to determine their
cross sections--which provide a major source of uncertainty in our consideration--one needs to
employ some models. In the present study, to estimate the ratio of the free total cross sections
for $K^+K^-$ production in $pn$ and $pp$ interactions, we have adopted the one-pion-exchange
model [57, 58]. It gives [58] for the ratio of the total cross sections of the reactions
$pp \to K^+{\bar K^0}pn$ and $pp \to K^0{\bar K^0}pp$ factor of 4, which is well in line with the
available experimental data. Evidently, this enables us to use the above model to estimate the ratio
between the free total cross sections of reactions (3) and (1) as well as to expect the obtained
ratio to be essentially correct and meaningful. Employing the ratios (8) between total cross sections
of different charge channels ${\pi}N \to K{\bar K}N$ found in [58] and accounting for that within the
one-pion-exchange model the reaction (1) is described by the four diagrams with the exchanged $\pi^0$
meson and the process (3) is also represented by the four diagrams, respectively, with the exchanged
$\pi^0$, $\pi^0$, $\pi^+$ and $\pi^-$ mesons, we get for this ratio the value of 4.5
\footnote{$^)$In this connection it should be noted that the ratio of the free total cross sections
of the reactions $pn \to pn{\phi} \to pnK^+K^-$ and $pp \to pp{\phi} \to ppK^+K^-$ is about 3--5
as follows from the results obtained in [59] within an effective meson-nucleon theory.}$^)$
.
Since the detailed
information on the reactions ${\pi}N \to K{\bar K}Nm{\pi}$ is not available, we shall not concern ourselves
here with this type of model to estimate the ratio of the total cross sections of channels (4) and (2) and
will assume that in the relevant energy region the cross section
$\sigma^{\rm nr}_{pn\to K^+K^-X}(\sqrt{s},\sqrt{s_{\rm th}})$ is related to
$\sigma^{\rm nr}_{pp\to K^+K^-X}(\sqrt{s},\sqrt{s_{\rm th}})$, defined above by equation (21), as:
\begin{equation}
\sigma^{\rm nr}_{pn\to K^+K^-X}(\sqrt{s},\sqrt{s_{\rm th}})=4.5
\sigma^{\rm nr}_{pp\to K^+K^-X}(\sqrt{s},\sqrt{s_{\rm th}}).
\end{equation}
The primary elementary cross sections (21) and (22) will be used below to calculate the yield of
$K^+K^-$ pairs from proton--nucleus collisions.

  Let us now define the exclusive differential cross section for the non-resonant $K^+K^-$ production
in $pA$ collisions from primary processes (1)--(4), corresponding
to the kinematical conditions of the ANKE experiment. In this experiment, the differential cross section
for production of $K^-$ mesons in the polar angular range of $0^{\circ} \le \theta_{K^-} \le 12^{\circ}$
in lab system in the interaction of protons of energy of 2.83 GeV with the C, Cu, Ag, and Au target nuclei
in coincidence with $K^+$ mesons, which were required to have vacuum momenta in the interval of
0.2 GeV/c $\le p_{K^+} \le$ 0.6 GeV/c and to be in the same polar angular domain in l.s. as that for
antikaons--$0^{\circ} \le \theta_{K^+} \le 12^{\circ}$, was measured as a function of their vacuum
momentum. Additionally, the invariant mass of the detected $K^+K^-$ pair, $IM(K^+K^-)$, was required
to be less than 1.005 GeV. Keeping this in mind and imposing the respective integrations on the full
exclusive differential cross section (9), we can represent this differential cross section in the
following form:
\begin{equation}
\frac{d\sigma_{pA\to K^+K^-X}^{({\rm prim})}
({\bf p}_0,AW,p_{K^-})}
{dp_{K^+}d{\bf \Omega}_{K^+}dp_{K^-}d{\bf \Omega}_{K^-}}=
\frac{1}{(2\pi)^2(1-\cos{12^{\circ}})^2~0.4~{\rm GeV/c}}
\end{equation}
$$
\times
\int\limits_{0.2~{\rm GeV/c}}^{0.6~{\rm GeV/c}}dp_{K^+}
\int\limits_{\cos{12^{\circ}}}^{1}d\cos{\theta_{K^+}}
\int\limits_{\cos{12^{\circ}}}^{1}d\cos{\theta_{K^-}}
\int\limits_{0}^{2\pi}d\phi_{K^+}
\int\limits_{0}^{2\pi}d\phi_{K^-}
$$
$$
\times
\frac{d\sigma_{pA\to K^+K^-X}^{({\rm prim})}
({\bf p}_0,{\bf p}_{K^+},{\bf p}_{K^-})}
{d{\bf p}_{K^+}d{\bf p}_{K^-}}p^2_{K^+}p^2_{K^-}
\theta[1.005~{\rm GeV}-IM(K^+K^-)],
$$
where
\begin{equation}
AW=0.2~{\rm GeV/c} \le p_{K^+} \le 0.6~{\rm GeV/c},\\\
0^{\circ} \le \theta_{K^{\pm}} \le 12^{\circ},\\\
IM(K^+K^-) \le 1.005~{\rm GeV};
\end{equation}
\begin{equation}
IM(K^+K^-)=\sqrt{(E_{K^+}+E_{K^-})^2-
({\bf p}_{K^+}+{\bf p}_{K^-})^2}.
\end{equation}
Here, $\phi_{K^+}$ and $\phi_{K^-}$ are the azimuthal angles of kaon and antikaon
momenta ${\bf p}_{K^+}$ and ${\bf p}_{K^-}$ in lab system and $\theta(x)$ is the standard
step function.

\section*{2.2. Two-step non-resonant $K^+K^-$ pair production mechanism}

\hspace{1.5cm} At the initial energy of our interest, 2.83 GeV, the following two-step processes
with pions in an intermediate states may contribute to the non-resonant kaon pair production
in $pA$ reactions:
\begin{equation}
p+N_1 \to \pi+X,
\end{equation}
\begin{equation}
\pi+N_2 \to N+K^++K^-.
\end{equation}
Here, $N_1$, $N_2$, $N$ and $\pi$ stand for $p$, $n$ and $\pi^+$, $\pi^0$, $\pi^-$ for the specific
isospin channel. Using the results given in [37, 43, 48] and accounting for the equation (9), the exclusive
differential non-resonant $K^+K^-$ production cross section for $pA$ collisions at small laboratory angles
from the secondary channels (27) can be represented as follows:
\begin{equation}
\frac{d\sigma_{pA\to K^+K^-X}^{({\rm sec})}
({\bf p}_0,{\bf p}_{K^+},{\bf p}_{K^-})}
{d{\bf p}_{K^+}d{\bf p}_{K^-}}=\frac{I^{\rm sec}_{V}[A]}{I^{\prime}_{V}[A]}
\sum_{\pi^{\prime}=\pi^+,\pi^0,\pi^-}\int \limits_{4\pi}d{\bf \Omega}_{\pi}
\int \limits_{p_{\pi}^{{\rm abs}}}^{p_{\pi}^{{\rm lim}}
(\vartheta_{\pi})}p_{\pi}^{2}
dp_{\pi}
\frac{d\sigma_{pA\to {\pi^{\prime}}X}^{({\rm prim})}({\bf p}_0)}{d{\bf p}_{\pi}}
\end{equation}
$$
\times
\left[\frac{Z}{A}\left<\frac{d\sigma^{\rm nr}_{{\pi^{\prime}}p \to NK^+K^-}({\bf p}_{\pi},
{\bf p}^{\prime}_{K^+},{\bf p}^{\prime}_{K^-})}{d{\bf p}^{\prime}_{K^+}d{\bf p}^{\prime}_{K^-}}\right>_A+
\frac{N}{A}\left<\frac{d\sigma^{\rm nr}_{{\pi^{\prime}}n \to NK^+K^-}({\bf p}_{\pi},
{\bf p}^{\prime}_{K^+},{\bf p}^{\prime}_{K^-})}{d{\bf p}^{\prime}_{K^+}d{\bf p}^{\prime}_{K^-}}\right>_A\right]
\frac{d{\bf p}^{\prime}_{K^+}}{d{\bf p}_{K^+}}
\frac{d{\bf p}^{\prime}_{K^-}}{d{\bf p}_{K^-}},
$$
where
\begin{equation}
I^{\rm sec}_{V}[A]=2{\pi}A^2\int\limits_{0}^{R}r_{\bot}dr_{\bot}
\int\limits_{-\sqrt{R^2-r_{\bot}^2}}^{\sqrt{R^2-r_{\bot}^2}}dz
\rho(\sqrt{r_{\bot}^2+z^2})
\int\limits_{0}^{\sqrt{R^2-r_{\bot}^2}-z}dl
\rho(\sqrt{r_{\bot}^2+(z+l)^2})
\end{equation}
$$
\times
\exp{\left[-\sigma_{pN}^{\rm in}A\int\limits_{-\sqrt{R^2-r_{\bot}^2}}^{z}\rho(\sqrt{r_{\bot}^2+x^2})dx
-\sigma_{{\pi^{\prime}}N}^{\rm tot}A\int\limits_{z}^{z+l}
\rho(\sqrt{r_{\bot}^2+x^2})dx\right]}
$$
$$
\times
\exp{\left[-\sigma_{K^+N}^{\rm tot}A\int\limits_{z+l}^{\sqrt{R^2-r_{\bot}^2}}
\rho(\sqrt{r_{\bot}^2+x^2})dx
-\int\limits_{z+l}^{\sqrt{R^2-r_{\bot}^2}}
\mu_{K^-N}\left[p^{\prime}_{K^-}(\sqrt{r_{\bot}^2+x^2})\right]
\rho(\sqrt{r_{\bot}^2+x^2})dx\right]}
$$
and
\begin{equation}
\left<\frac{d\sigma^{\rm nr}_{{\pi^{\prime}}N \to NK^+K^-}({\bf p}_{\pi},
{\bf p}^{\prime}_{K^+},{\bf p}^{\prime}_{K^-})}{d{\bf p}^{\prime}_{K^+}d{\bf p}^{\prime}_{K^-}}\right>_A
=
\int\int
P_A({\bf p}_t,E)d{\bf p}_tdE
\end{equation}
$$
\times
\left\{\frac{d\sigma^{\rm nr}_{{\pi^{\prime}}N \to NK^+K^-}[\sqrt{s_1},<m^*_{K^+}>,<m^*_{K^-}>,<m^*_{N}>,
{\bf p}^{\prime}_{K^+},{\bf p}^{\prime}_{K^-}]}{d{\bf p}^{\prime}_{K^+}d{\bf p}^{\prime}_{K^-}}\right\}.
$$
Here,
$d\sigma^{\rm nr}_{{\pi^{\prime}}N \to NK^+K^-}[\sqrt{s_1},<m^*_{K^+}>,<m^*_{K^-}>,<m^*_{N}>,
{\bf p}^{\prime}_{K^+},{\bf p}^{\prime}_{K^-}]/d{\bf p}^{\prime}_{K^+}d{\bf p}^{\prime}_{K^-}$
are the "in-medium" exclusive differential cross sections for the non-resonant production of $K^+$ and $K^-$
mesons with the in-medium momenta ${\bf p}^{\prime}_{K^+}$ and ${\bf p}^{\prime}_{K^-}$,
respectively, in reactions (27) at the ${\pi}N$ centre-of-mass energy $\sqrt{s_1}$ and the other quantities,
entering into the equations (28) and (29), are defined in [37, 43, 48].

    The elementary $K^+K^-$ creation processes ${\pi^+}n \to pK^+K^-$, ${\pi^0}p \to pK^+K^-$,
${\pi^0}n \to nK^+K^-$ and ${\pi^-}p \to nK^+K^-$ have been included in our calculations of the
non-resonant kaon pair production on nuclei. In them, the exclusive differential cross sections
$d\sigma^{\rm nr}_{{\pi^{\prime}}N \to NK^+K^-}[\sqrt{s_1},<m^*_{K^+}>,<m^*_{K^-}>,<m^*_{N}>,
{\bf p}^{\prime}_{K^+},{\bf p}^{\prime}_{K^-}]/d{\bf p}^{\prime}_{K^+}d{\bf p}^{\prime}_{K^-}$
have been computed according to the three-body phase space [43] with accounting for the medium effects
on the outgoing in reactions (27) nucleons, kaons and antikaons on the same footing as that employed
in calculating the $K^+K^-$ production cross sections (16) from primary proton-induced reaction
channels:
\begin{equation}
\frac{d\sigma^{\rm nr}_{{\pi^{\prime}}N \to NK^+K^-}[\sqrt{s_1},<m^*_{K^+}>,<m^*_{K^-}>,<m^*_{N}>,
{\bf p}^{\prime}_{K^+},{\bf p}^{\prime}_{K^-}]}{d{\bf p}^{\prime}_{K^+}d{\bf p}^{\prime}_{K^-}}=
\frac{\sigma^{\rm nr}_{{\pi^{\prime}}N \to NK^+K^-}(\sqrt{s_1},\sqrt{s^*_{1,{\rm th}}})}
{8E^{\prime}_{K^+}E^{\prime}_{K^-}}
\end{equation}
$$
\times
\frac{1}{I_3(s_1,<m^*_{K^+}>,<m^*_{K^-}>,<m^*_N>)}
\frac{1}{(\omega_1+E_t)}\delta\left[\omega_1+E_t-\sqrt{(<m^*_N>)^2+({\bf Q}_1+{\bf p}_t)^2}\right],
$$
where
\begin{equation}
\omega_1=E_{\pi}-E^{\prime}_{K^+}-E^{\prime}_{K^-},\\\
{\bf Q}_1={\bf p}_{\pi}-{\bf p}^{\prime}_{K^+}-{\bf p}^{\prime}_{K^-},
\end{equation}
${\bf p}_{\pi}$ and $E_{\pi}$ are the momentum and total energy of an intermediate pion
(which is assumed to be on-shell), $I_3$ is the three-body phase space defined in [43]
by the formula (27).
In equation (31), $\sigma^{\rm nr}_{{\pi^{\prime}}N \to NK^+K^-}(\sqrt{s_1},\sqrt{s^*_{1,{\rm th}}})$
are the "in-medium"
total cross sections for the non-resonant $K^+K^-$ pair production in reactions (27) having the
threshold energy $\sqrt{s^*_{1,{\rm th}}}=<m^*_{K^+}>+<m^*_{K^-}>+<m^*_N>$. As before, we assume
that these cross sections are equivalent to the vacuum ones
$\sigma^{\rm nr}_{{\pi^{\prime}}N \to NK^+K^-}(\sqrt{s_1},\sqrt{s_{1,{\rm th}}})$
in which the free threshold energy $\sqrt{s_{1,{\rm th}}}=2m_K+m_N$
is replaced by the in-medium threshold $\sqrt{s^*_{1,{\rm th}}}$.
In line with the equation (20), for the free non-resonant total cross sections
$\sigma^{\rm nr}_{{\pi^{\prime}}N \to NK^+K^-}(\sqrt{s_1},\sqrt{s_{1,{\rm th}}})$
we have adopted the following expression:
\begin{equation}
\sigma^{\rm nr}_{{\pi^{\prime}}N \to NK^+K^-}(\sqrt{s_1},\sqrt{s_{1,{\rm th}}})=
\sigma_{{\pi^{\prime}}N \to NK^+K^-}(\sqrt{s_1},\sqrt{s_{1,{\rm th}}})-
\sigma_{{\pi^{\prime}}N \to N\phi}(\sqrt{s_1})BR(\phi \to K^+K^-),
\end{equation}
where $\sigma_{{\pi^{\prime}}N \to NK^+K^-}(\sqrt{s_1},\sqrt{s_{1,{\rm th}}})$ are
the corresponding total cross sections
which include the contributions from $\phi$ and non-$\phi$ components, whereas
$\sigma_{{\pi^{\prime}}N \to N\phi}(\sqrt{s_1})$ are the total cross sections
of the ${\pi^{\prime}}N \to N\phi$ reactions.
For the total cross sections
$\sigma_{{\pi}^{\prime}N \to NK^+K^-}(\sqrt{s_1},\sqrt{s_{1,{\rm th}}})$
and $\sigma_{{\pi}^{\prime}N \to N\phi}(\sqrt{s_1})$
we have used in our model calculations the parametrizations (63) and (13)
suggested in [43] and [60], respectively.

  Before closing this subsection, one needs to define the exclusive differential cross section
for the non-resonant $K^+K^-$ production in $pA$ interactions from the secondary processes (27),
corresponding to the kinematical conditions of the ANKE experiment. Analogously to (23), the latter
cross section can be determined as:
\begin{equation}
\frac{d\sigma_{pA\to K^+K^-X}^{({\rm sec})}
({\bf p}_0,AW,p_{K^-})}
{dp_{K^+}d{\bf \Omega}_{K^+}dp_{K^-}d{\bf \Omega}_{K^-}}=
\frac{1}{(2\pi)^2(1-\cos{12^{\circ}})^2~0.4~{\rm GeV/c}}
\end{equation}
$$
\times
\int\limits_{0.2~{\rm GeV/c}}^{0.6~{\rm GeV/c}}dp_{K^+}
\int\limits_{\cos{12^{\circ}}}^{1}d\cos{\theta_{K^+}}
\int\limits_{\cos{12^{\circ}}}^{1}d\cos{\theta_{K^-}}
\int\limits_{0}^{2\pi}d\phi_{K^+}
\int\limits_{0}^{2\pi}d\phi_{K^-}
$$
$$
\times
\frac{d\sigma_{pA\to K^+K^-X}^{({\rm sec})}
({\bf p}_0,{\bf p}_{K^+},{\bf p}_{K^-})}
{d{\bf p}_{K^+}d{\bf p}_{K^-}}p^2_{K^+}p^2_{K^-}
\theta[1.005~{\rm GeV}-IM(K^+K^-)],
$$
where $AW$ denotes ANKE acceptance window and it is defined above by the relations (24).

  Let us discuss now the results of our calculations in the framework of the approach outlined above.

\section*{3. Results}

\hspace{1.5cm} At first, we consider the non-resonant differential $K^+K^-$ production cross sections
in the ANKE acceptance window from the one-step and two-step creation mechanisms in $p$C and $p$Au
collisions calculated on the basis of equations (23) and (34) for the proton kinetic energy of
2.83 GeV in two scenarios for the $K^-$ potential depth $U^0_{K^-}$, namely: i) $U^0_{K^-}=0$ MeV and
ii) $U^0_{K^-}=-126$ MeV. In our calculations of these cross sections, presented in figure 2,
the effective nuclear potentials $U^0_{K^+}$, $U^0_{N}$ as well as kaon,
antikaon and proton Coulomb potentials, entering into the equation (5) determining the average in-medium
hadron mass, were chosen equal to those introduced before and were retained to be fixed throughout the
subsequent calculations reported below.
\begin{figure}[htb]
\begin{center}
\includegraphics[width=12.0cm]{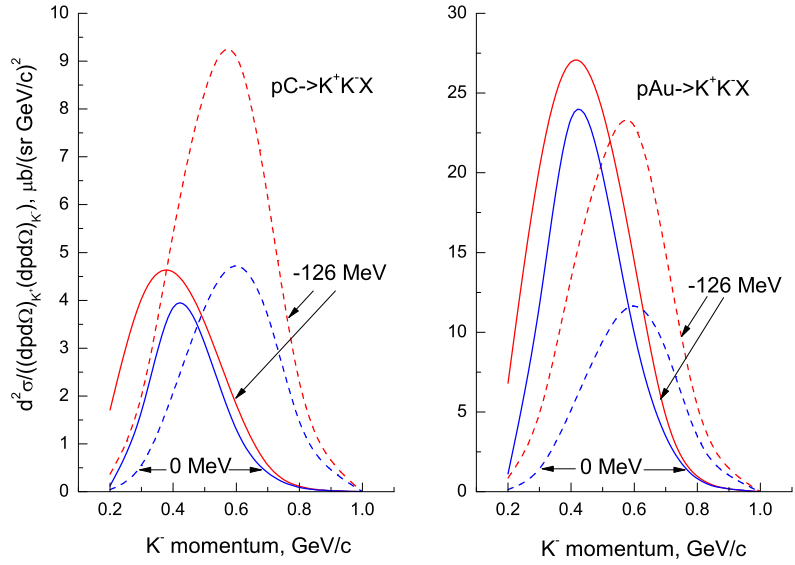}
\vspace*{-2mm} \caption{(color online) Exclusive differential cross section for
the non-resonant production of $K^+K^-$ pairs
from primary and secondary channels (dashed and solid lines, respectively)
in the ANKE acceptance window as a function of antikaon momentum
in the interaction of protons of energy of 2.83 GeV with C (left panel)
and Au (right panel) nuclei for $K^-$ potential depths $U^0_{K^-}=0$ MeV and
$U^0_{K^-}=-126$ MeV as indicated by the curves.}
\label{void}
\end{center}
\end{figure}
One can see that the two-step $K^+K^-$ production mechanism is of importance compared to the one-step
in the chosen kinematics at laboratory antikaon momenta $\le$ 0.4 GeV/c and $\le$ 0.6 GeV/c for target
nuclei C and Au, respectively, for both considered options for the $K^-$ potential depth $U^0_{K^-}$,
whereas at higher $K^-$ momenta the direct $K^+K^-$ production processes (1)--(4) are dominant and their
dominance here is more pronounced, as is expected, for the carbon target nucleus. This indicates that the
secondary pion--nucleon $K^+K^-$ production channels have to be taken into account in the analysis of the
data on non-resonant kaon pair creation in $pA$ interactions obtained recently by
the ANKE-at-COSY Collaboration. It is also clearly seen that the considered coincident antikaon spectrum
from the one-step production mechanism reacts sensitively to the $K^-$ potential practically at all
antikaon momenta involved. The dependence of the $K^-$ yield from the two-step production
mechanism on this potential exists as well, but, contrary to the preceding case, it is more moderate.
Therefore, as a result, the above offers the possibility to determine the antikaon potential experimentally
(cf. figures 4, 5 given below).
\begin{figure}[!h]
\begin{center}
\includegraphics[width=12.0cm]{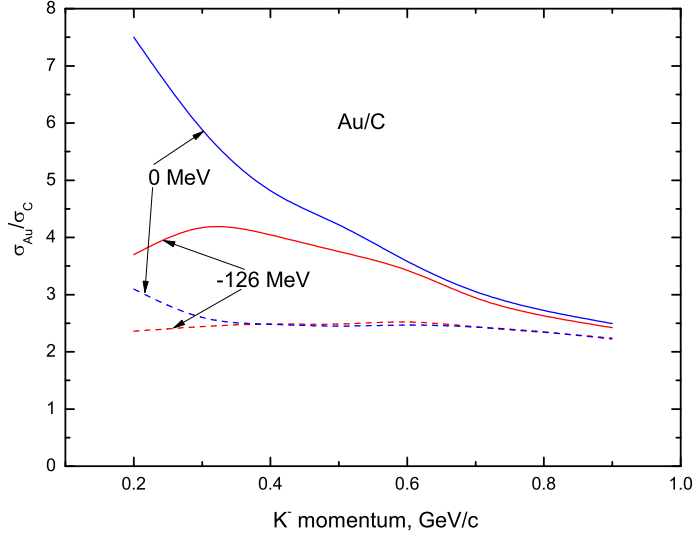}
\vspace*{-2mm} \caption{(color online) Ratio of the exclusive differential cross section
for the non-resonant production of $K^+K^-$ pairs
off Au target nucleus in the ANKE acceptance window to that off C target nucleus as a function of antikaon momentum
for the one-step and one--plus two-step $K^+K^-$ production mechanisms (dashed and solid lines, respectively)
for incident energy of 2.83 GeV as well as
for $K^-$ potential depths $U^0_{K^-}=0$ MeV and $U^0_{K^-}=-126$ MeV as indicated by the curves.}
\label{void}
\end{center}
\end{figure}
\begin{figure}[!h]
\begin{center}
\includegraphics[width=12.0cm]{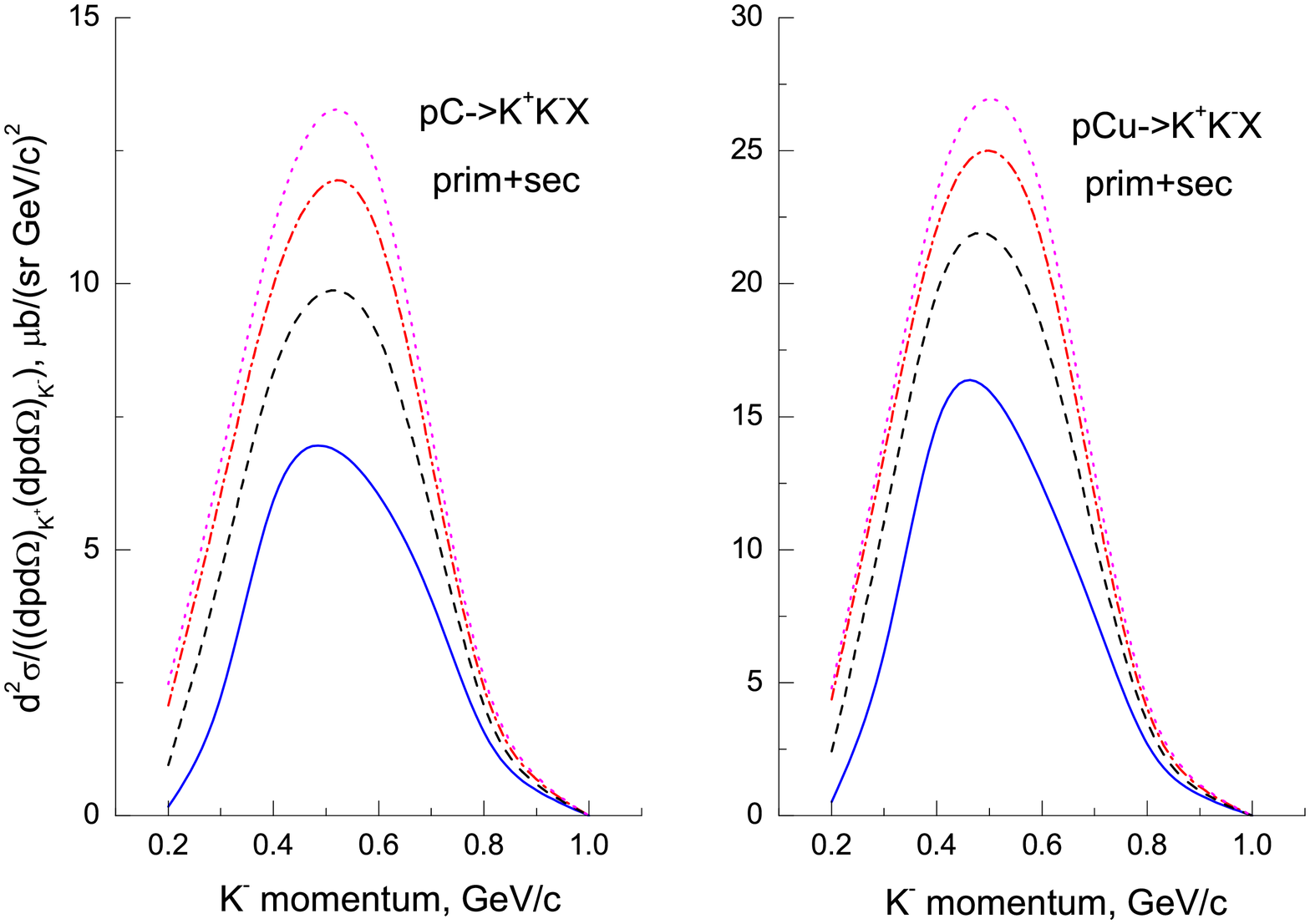}
\vspace*{-2mm} \caption{(color online) Exclusive differential cross section for
the non-resonant production of $K^+K^-$ pairs
from primary plus secondary channels
in the ANKE acceptance window as a function of antikaon momentum
in the interaction of protons of energy of 2.83 GeV with C (left panel)
and Cu (right panel) target nuclei
for $K^-$ potential depths $U^0_{K^-}=0$ MeV (solid lines), $U^0_{K^-}=-60$ MeV (dashed lines),
$U^0_{K^-}=-126$ MeV (dot-dashed lines) and $U^0_{K^-}=-180$ MeV (dotted lines).}
\label{void}
\end{center}
\end{figure}
\begin{figure}[!h]
\begin{center}
\includegraphics[width=12.0cm]{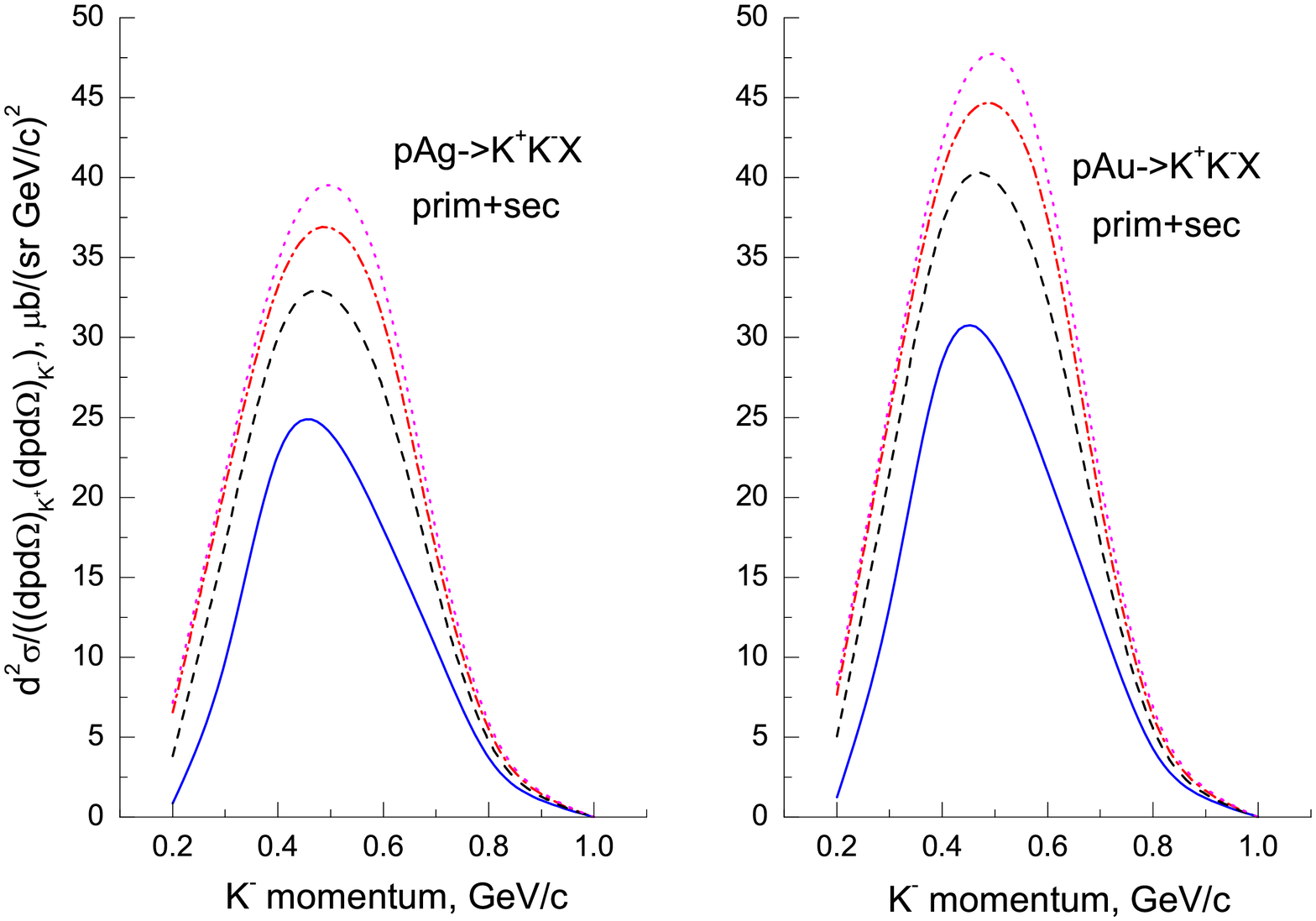}
\vspace*{-2mm} \caption{(color online) The same as in figure 4, but for the Ag and Au target nuclei.}
\label{void}
\end{center}
\end{figure}

In figure 3 we show our predictions for the another observable--the ratio of the exclusive differential
cross section for the non-resonant production of $K^+K^-$ pairs off Au target nucleus in the ANKE acceptance
window to that off carbon target nucleus as a function of antikaon momentum for the one-step and
one- and two-step kaon pair creation mechanisms for the projectile energy of 2.83 GeV and within the considered
above two scenarios for the $K^-$ potential depth $U^0_{K^-}$
\footnote{$^)$It should be noticed that the similar inclusive observable was used in [8] and [11]
in extracting the $K^+$ and $K^0$ nuclear potentials by comparing the inclusive data with the
model calculations. Such relative observables are more favorable compared to those based on the
absolute cross sections for the aim of getting the information on particle nuclear potential
both from the experimental and theoretical sides, since, on the one hand,
they allow for a reduction of systematic errors due to the cancellation of the efficiency corrections
and, on the other hand, the theoretical uncertainties associated with the particle production and
absorption mechanisms substantially cancel out in them.}$^)$
.
It can be seen that there are essential differences between the results obtained by using different suppositions
about the $K^+K^-$ creation mechanism and the same potential depths $U^0_{K^-}$
(between solid and dashed lines). We may see, for example, that for low
antikaon momenta, where the $K^+K^-$ production via the secondary pion-induced reaction channels is enhanced,
the calculated ratio can be of the order of 2.4 and 3.1 for the direct $K^+K^-$ creation mechanism as well as
3.7 and 7.5 for the direct plus two-step $K^+K^-$ production mechanisms when the antikaon potential depths of
-126 MeV and 0 MeV were applied, respectively. Therefore, we can conclude that in the analysis of data on this
relative observable, taken in the ANKE experiment, it is important to account for the secondary
pion-induced $K^+K^-$ production processes. Looking at this figure, one can see also that the calculated ratio
for the direct $K^+K^-$ production mechanism, contrary to the absolute cross sections depicted in figure 2,
is not very sensitive to the effective potential $U^0_{K^-}$, which is seen inside the nucleus by an antikaon,
practically at all $K^-$ momenta of interest. Whereas, this potential influences substantially the ratio of
the differential cross sections under consideration obtained with accounting for the one- and two-step creation
mechanisms, namely: the inclusion of the attractive $K^-$ potential leads to the visible suppression of this
ratio at low antikaon momenta (cf. figure 6 given below). This leaves a room for distinguishing between
different $K^-$ potentials from the measurements of the relative observable like that just considered.
\begin{figure}[!h]
\begin{center}
\includegraphics[width=16.0cm]{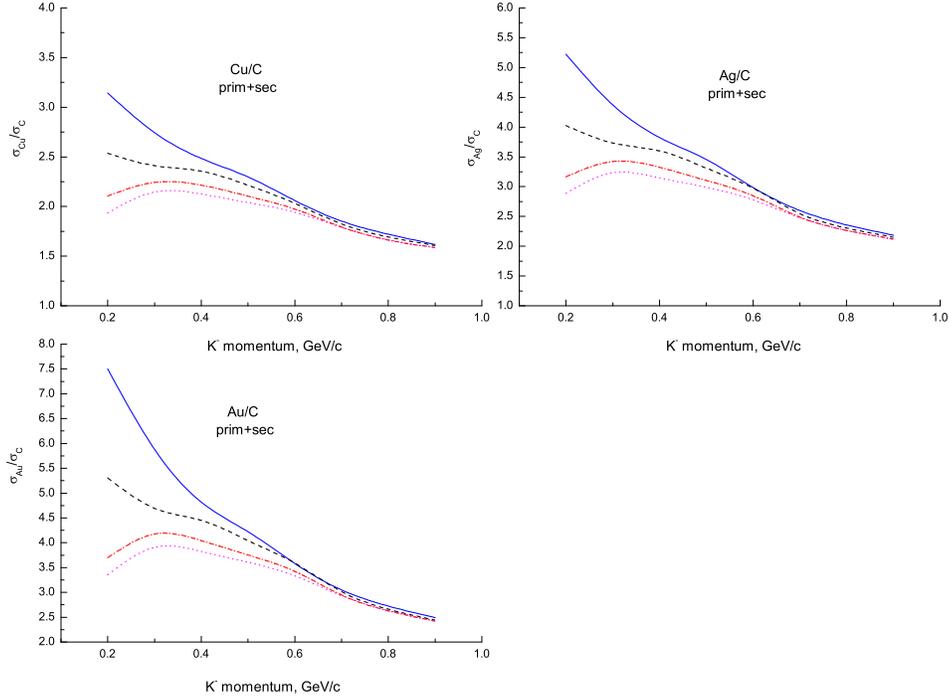}
\vspace*{-2mm} \caption{(color online) Ratio of the exclusive differential cross section for the
non-resonant production of $K^+K^-$ pairs
off Cu, Ag and Au target nuclei presented in figures 4, 5 to that off C target nucleus given in figure 4
as a function of antikaon momentum
for $K^-$ potential depths $U^0_{K^-}=0$ MeV (solid lines), $U^0_{K^-}=-60$ MeV (dashed lines),
$U^0_{K^-}=-126$ MeV (dot-dashed lines) and $U^0_{K^-}=-180$ MeV (dotted lines).}
\label{void}
\end{center}
\end{figure}

    Figures 4 and 5 show the results of our overall model calculations following the
equations (23) and (34) for the
exclusive differential cross sections for the non-resonant production of $K^+K^-$ pairs
on C, Cu and Ag, Au target nuclei in the ANKE acceptance window for the primary and secondary
$K^+K^-$ creation processes obtained for bombarding energy of 2.83 GeV by employing in them
four adopted options for the $K^-$ potential depth $U^0_{K^-}$ to see more clearly the sensitivity
of the calculated cross sections to the choice of the antikaon optical potential. It is nicely seen
that for all nuclei there are measurable changes in the absolute cross sections for the $K^+K^-$
production on these nuclei due to this potential, especially when it varies between -60 MeV and
0 MeV (cf. figure 2). We, therefore, can conclude that the $K^-$ nuclear potential can be in
principle extracted from the direct comparison of the results of our calculations, presented in
figures 4 and 5, with the data from the ANKE experiment.

    In figure 6 we show also the overall predictions of our model for the
antikaon momentum dependence of the ratios of the exclusive differential cross sections for the
non-resonant production of $K^+K^-$ pairs off Cu, Ag, and Au target nuclei, given, respectively,
in figures 4 and 5, to that off C target nucleus, presented in figure 4, for four adopted options
for the $K^-$ potential depth $U^0_{K^-}$. One can see that the relative $K^+K^-$ yield, along
with the absolute one considered before, is appreciably sensitive to the antikaon potential at
momenta less than 0.6 GeV/c for all considered $A$/C combinations, which is consistent with our
previous findings of figure 3. Thus, our results demonstrate that the measurements of the $K^-$
momentum dependence of the relative cross section for non-resonant $K^+K^-$ production in $pA$
collisions in the chosen kinematics and at beam energy of interest will also allow one to discriminate
between four employed scenarios for the in-medium antikaon modification.
\begin{figure}[!h]
\begin{center}
\includegraphics[width=12.0cm]{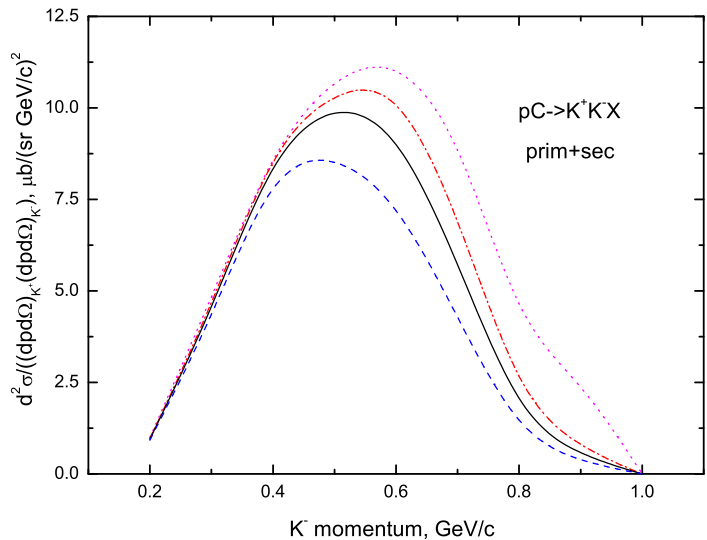}
\vspace*{-2mm} \caption{(color online) Exclusive differential cross section for
the non-resonant production of $K^+K^-$ pairs from primary plus secondary channels
in the laboratory polar angular acceptance window $0^{\circ} \le \theta_{K^{\pm}} \le \theta_{max}$
as a function of antikaon momentum in the interaction of protons of energy of 2.83 GeV with C
target nucleus for $K^-$ potential depth $U^0_{K^-}=-60$ MeV and for
$\theta_{max}=5^{\circ}$ (dotted line), $\theta_{max}=10^{\circ}$ (dot-dashed line),
$\theta_{max}=12^{\circ}$ (solid line), and $\theta_{max}=15^{\circ}$ (dashed line).}
\label{void}
\end{center}
\end{figure}

To extend the potential range of applicability of our model we performed the calculations of the
exclusive differential cross section for the non-resonant production of $K^+K^-$ pairs
from primary plus secondary channels for a carbon target nucleus in the scenario with $K^-$ potential
depth of -60 MeV at saturation density and for different polar angular acceptance windows
for $K^+$ and $K^-$ mesons. They are
shown in figure 7. Here one can see a reduction of the cross section with respect to the case of
$\theta_{max}=5^{\circ}$ with increasing the maximal angle of acceptance window $\theta_{max}$
at antikaon momenta larger than 0.4 GeV/c
\footnote{$^)$For example, the cross section is decreased by a factors of about 2 and 2.5 at momenta
$\sim$ 0.7--0.8 GeV/c when going from $\theta_{max}=5^{\circ}$ to $\theta_{max}=12^{\circ}$ and to
$\theta_{max}=15^{\circ}$, respectively.}$^)$
,
whereas it is practically not changed at lower $K^-$ momenta.
Such behavior of the overall cross section of interest with increasing the $K^{\pm}$ creation polar angular
bin is expected to be observed both for other adopted options for antikaon potential depth $U^0_{K^-}$
and for other target nuclei considered in the present paper.
It can be explained by the following. At low $K^-$ momenta, the double differential cross sections,
appearing in the equations (23) and (34), after averaging over the studied range of kaon momenta with
accounting for the kinematical constraint $IM(K^+K^-) \le 1.005~{\rm GeV}$ depend weakly on the $K^{\pm}$
production angles $\theta_{K^{\pm}}$ when they vary close to zero angle. Therefore, the integrals in
formulas (23) and (34) are practically proportional to the solid angle, covered by the acceptance window,
squared. Due to the employed definition of cross sections (23) and (34), this leads to weak
dependence of the overall exclusive cross section defined by these cross sections on the size of the acceptance
window. At high $K^-$ momenta, the reduction of the cross section with increasing this size is caused by
the strong decrease of the above double differential cross sections with enlarging the $K^{\pm}$
production angles $\theta_{K^{\pm}}$.
In view of the expected data from ANKE experiment, the results presented in figure 7 can be also used as
an additional tool to those given before for determining the $K^-$ mean-field nuclear potential at studied
range of antikaon momenta.

   Taking into account the above considerations, we come to the conclusion that the absolute and relative
observables such as the exclusive $K^-$ momentum distribution and the ratio of this distribution on heavy
nucleus to that on light one (C), considered in the present work, can be useful to help determine
the anikaon-nucleus optical potential.

\section*{4. Conclusions}

\hspace{1.5cm} In this paper we calculated the antikaon momentum dependences of the exclusive
absolute and relative non-resonant $K^+K^-$ pair yields from
$pA$ ($A$=C, Cu, Ag, and Au) collisions at 2.83 GeV beam energy
in the acceptance window of the ANKE spectrometer, used in a recent experiment performed at COSY,
by considering incoherent primary proton--nucleon and secondary pion--nucleon kaon pair
production processes in the framework of a nuclear spectral function approach, which accounts for,
in particular, nuclear mean-field potential effects on these processes.
It was found that these observables are strongly sensitive to the antikaon-nucleus optical potential
in the studied region of the antikaon momenta.
This gives a nice opportunity to determine it from the direct comparison of the results
of our calculations with the upcoming data from the respective ANKE-at-COSY experiment.
It was also shown that the pion-nucleon production channels dominate in the low-momentum $K^-$, $K^+$
production in the considered kinematics and, therefore, they should be accounted for in the analysis
of these data with the aim to obtain information on the antikaon mean-field nuclear potential.
\\
\\
{\bf Acknowledgments}
\\
The authors gratefully acknowledge A. Polyanskiy and H. Str$\ddot{\rm o}$her
for their interest in this work.
\\

\end{document}